\documentclass[12pt]{article}
\usepackage{epsfig}

\newcommand{\boldtau}{\mbox{\boldmath $\tau$}}
\newcommand{\boldphi}{\mbox{\boldmath $\phi$}}

\begin{document}

\title{ Effective field theory contribution to charge 
symmetry breaking $NN$ scattering }

\author{J. A.
Niskanen{\thanks{Email: jouni.niskanen@helsinki.fi}}\\
{\it Department of Physics, P. O. Box 64} \\
{\it FIN-00014 University of Helsinki, Finland}}


\maketitle
\begin{abstract}
The effect of isospin breaking pion $s$-wave
rescattering is included in elastic $NN$ scattering
at low energies using effective field theory.  
Although this mechanism gave a large contribution
to charge symmetry breaking
 in $np\rightarrow d\pi^0$, the effect is rather
small in $pp$ vs. $nn$ scattering parameters and in
the $^3$H-$^3$He binding energy difference. This smallness 
is caused by large cancellation of the up-down quark
mass difference contribution and electromagnetic effects 
to the $np$ mass difference.

\end{abstract}



Charge symmetry is the most accurate special case of general flavour
symmetry.
It is trivially broken by the electromagnetic interaction, notably the
Coulomb force in comparisons of the $pp$ and $nn$ systems and by
the magnetic interaction in the $np$ system. Other
well known sources are the $np$ mass difference and $\eta\pi$- as well
as $\rho\omega$-meson mixing. These in turn may be related to the
up- and down-quark mass difference - the microscopic flavour symmetry
breaking in QCD. One might consider remarkable the fact that, although
the relative quark mass difference is large ($\ge 10\%$), the symmetry
breaking at the observable hadron level is two orders of magnitude 
smaller.

Charge symmetry breaking (CSB) has been studied for the mirror system
$pp$ vs. $nn$ for many decades \cite{physrep},
while its appearance in the $np$ system was first seen only a decade
ago \cite{classiv} as the difference $\Delta A = A_n - A_p$ elastic
analyzing powers and is presently
being searched for also in pionic inelasticity in the reaction
$np \rightarrow d\pi^0$ \cite{e704}. The CSB observables 
have been seen in calculations
to be sensitive to different combinations of sources. For example, in
$np$ scattering above 300 MeV the $np$ mass difference in OPE dominates,
while at $\approx$ 200 MeV $\rho\omega$ meson mixing and the magnetic
interaction become about equally important\cite{wtm}. 
Of traditional
CSB mechanisms in pion production $\eta\pi$
mixing is important and was seen to dominate at threshold \cite{few},
while at higher energies the $np$ mass difference becomes more 
important\cite{nst}.
The CSB effects in the $np$ system change the isospin of the two 
baryons
(class IV in the terminology of Ref. \cite{hm}), while in
$pp$ and $nn$ the isospin is conserved (class III). In class III
the main contribution is expected to be the $\rho\omega$ meson
mixing \cite{physrep,cb}.

Two-meson exchange in CSB has been studied earlier extensively by
Coon and collaborators \cite{cs,cn} and in charge dependence
in e.g. Refs. \cite{cs,em,cm,fk}.

Recently a new mechanism related to the $ud$ quark mass 
difference in QCD based effective field
theory was suggested for the CSB forward-backward asymmetry of the
cross section in $np \rightarrow d\pi^0$ \cite{knm}. It consists
of CSB $s$-wave
rescattering of the pion from the second nucleon. This rescattering
(depicted in Fig. 1a) appears in effective field theory
through the second term of
the isospin symmetry violating Lagrangian \cite{kolck,weinberg}
\begin{equation}
 {\cal{L}} =
 \frac{\delta m_{N}}{2}  \left(N^\dagger \tau_{0}N
   -\frac{2}{DF_{\pi}^{2}}  N^\dagger \phi_{0}
\boldphi \cdot \boldtau N \right) ,
\end{equation}
where the nucleon isospin is represented by the Pauli matrices
$\boldtau$, $F_\pi=186$ MeV is the pion decay constant and
$\delta m_N$ is the up and down quark mass difference effect in the
nuclear masses. The
denominator is in principle $D=1+\boldphi^2/F_\pi^2$,
but $D=1$ is used here. The isospin violation here
originates from the rather significant quark mass difference
$m_d -m_u \approx $ a few MeV. In addition to the bare quark
mass difference one should include an electromagnetic
contribution  $\bar{\delta} m_{N}$ to the nucleon mass
difference changing the effective CSB strength parameter 
\cite{knm}. We shall come to this correction later.

 \begin{figure}[tb]
\begin{center}
\epsfig{figure=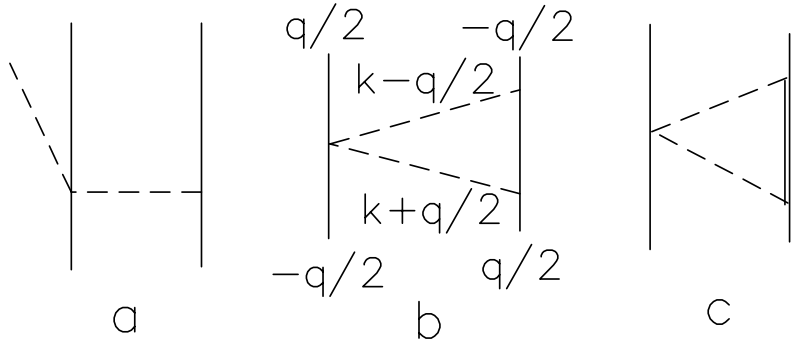,height=11cm,angle=90}
\end{center}
 \vspace{-.5cm}
\caption{CSB mechanisms arising from the up-down quark mass
difference in pion rescattering:
in $np \rightarrow d\pi^0$ (a), in $NN$ elastic scattering 
with a nucleonic (b) and $\Delta$ intermediate state (c).}
 \label{diags}
 \end{figure}

The above mechanism  was seen to be a major contributor
to the asymmetry in CSB pion production. However, it is clear
that returning the emitted pion back to the first nucleon
it can also contribute to elastic scattering as shown
in Fig. 1b and 1c. The question is only whether its
contribution really is isospin violating and of what type.
The aim of this paper is to investigate this interaction and
its effect to the difference of the $^1S_0$ scattering lengths
(experimentally estimated to be
$\Delta a = a_{pp} - a_{nn} = 1.5 \pm 0.5$ fm).
Furthermore, a simple estimate of this effect to the
$^3$H-$^3$He binding energy difference is made.

It is straightforward algebra to see that, with the conventions of
Fig. 1 and neglecting the baryon kinetic energies,
the diagram 1b yields in the momentum
space a CSB interaction of the form
\begin{equation}
\displaystyle
V_N(q) =
 \frac{\delta m_{N}}{F_\pi^2} \frac{f^2}{\mu^2}
\int \frac{d^3k}{(2\pi)^3} \frac{(k^2-q^2/4)(\tau_{10}+\tau_{20})}
{[\mu^2+({\bf k} +{\bf q}/2)^2] [\mu^2+({\bf k} -{\bf q}/2)^2]},
\end{equation}
where $f^2/4\pi = 0.076$ is the pion-nucleon coupling constant, $\mu$
the pion mass and $\tau_{i0}$ refers to the $z$ component of the
isospin
operator of the $i$th nucleon. With the intermediate $\Delta$
(Fig. 1c) the corresponding result would be
\begin{equation}
\displaystyle
V_\Delta(q) = \frac 4 9
 \frac{\delta m_{N}}{F_\pi^2}\,
 \frac{f^{*2}}{\mu^2}
\int \!\frac{d^3k}{(2\pi)^3} \frac{(k^2-q^2/4)(\tau_{10}+\tau_{20})
(\omega_+ +\omega_- + \Delta) }
{\omega_+ \omega_- (\omega_+ + \Delta) (\omega_- +\Delta)
(\omega_+ + \omega_-) },
\end{equation}
where now the $\pi N\Delta$ coupling constant is
${f^*}^2/4\pi = 0.35$ from the width of the $\Delta(1232)$ and
$\Delta$ is the mass difference between the $\Delta(1232)$ isobar
and the nucleon (the real part of the $\Delta$ pole is used).
Also a shorthand notation has been introduced for the pion energy 
with $\omega_\pm^2 = \mu^2+({\bf k} \pm{\bf q}/2)^2$.
In addition, monopole form factors $(\Lambda^2 - \mu^2)/(\Lambda^2
+ q_\pi^2)$ are inserted for the pion
emission and absorption vertices.
 Clearly the above potentials belong to class III in the
classification of \cite{hm}, which violates charge symmetry
between $pp$ and $nn$ but not in the $np$ system.\footnote{
One might note that there is also a contribution with a structure
$i({\mbox{\boldmath $\tau$}}_1 \pm {\mbox{\boldmath $\tau$}}_2)\,
 ({\mbox{\boldmath $\sigma$}}_1\pm{\mbox{\boldmath $\sigma$}}_2)
\cdot {\bf k}\times{\bf q}$.
With the above static approximation for the baryons this vanishes in
the integration over ${\bf k}$. However, if the baryon kinetic
energies are taken into account, there is also an odd term
in the angular dependence of the denominators allowing a
nonzero class IV part as found in Ref. \cite{tpe}. At low
energies this correction, however, should be significantly smaller
than the potentials (2-3).} For positive $\delta m_N$ 
they tend to make the $nn$ interaction more attractive.

An interesting point in these CSB contributions is that the
coefficient multiplying the integrals could be numerically large as
compared with the coefficients in Refs. \cite{cn,tpe} for CSB arising
from the $np$ mass difference. However, the dimension is different
(depending also on the integral). One may ask whether the
contribution could be even unrealistically large to exclude this
mechanism from CSB. An explicit calculation is necessary to answer
this question.

One may note that there is large uncertainty in the exact
value of $\delta m_N$ with estimates ranging mostly between 
2 and 3 MeV depending on electromagnetic corrections to the
$np$ mass difference. 
For the moment the value $\delta m_N = 2.4$ MeV has been used
in these results, which represented the total CSB strength
 for the reaction $np --> d\pi^0$ in Ref. \cite{knm}
(including also the electromagnetic contribution to the $np$
mass difference).
The contribution to $\Delta a $  scales linearly with
$\delta m_N$. We shall return to the effect
of the electromagnetic corrections later.

\begin{figure}[tb]
\begin{center}
\epsfig{figure=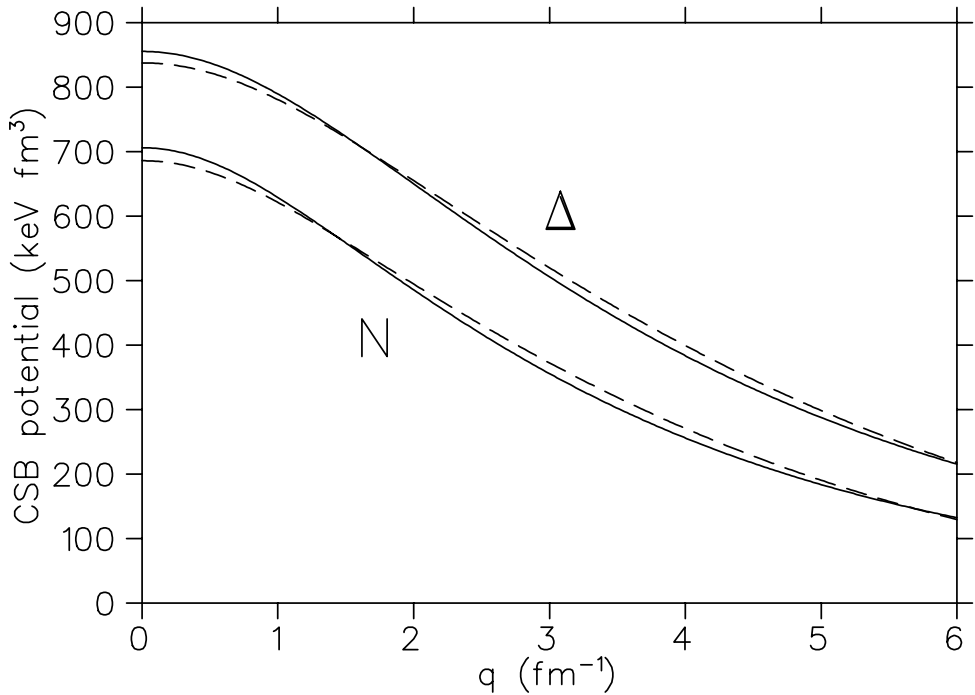,height=13cm,angle=90}
\end{center}
\vspace{-.5cm}
 \caption{The momentum space CSB potentials $V_N(q)$ and
$V_\Delta (q)$ as functions of the momentum transfer $q$.
 Solid curves: exact integrals used; dashed
curves: fits with forms (4-5).}
 \label{vq}
 \end{figure}

The above integrals are numerically easy to perform and,
in the same way as in Ref. \cite{tpe},
the resulting potential is then transformed
into the coordinate space where the final calculations are done.
Simple fits of the integrals with a form
\begin{equation}
V(q) = A \frac{B^2}{B^2+q^2}
\end{equation}
for the first ($k^2$ dependent) parts and
\begin{equation}
V(q) = A \frac{B^2}{B^2+q^2}\frac{C^2}{C^2+q^2}
\end{equation}
lead to a tolerable agreement (although not as perfect as in Ref.
\cite{tpe}) with the exact results for $V_N$ and $V_\Delta$
(Fig. 2). In the coordinate representation these turn to
Yukawa functions or their derivatives, shown in Fig. 3.
These are very large potentials, indeed, 
for charge asymmetry, an order
of magnitude larger than in Ref. \cite{tpe} for class IV, but
this may be in line with chiral power counting arguments,
which stipulate that class III should be stronger than class IV
\cite{fk,kolck}.
In these figures the coefficients of the $(\tau_{10} +
\tau_{20})$ operators are shown, so the {\it total difference} of
the $pp$ vs. $nn$ interaction will get still another factor of 4.

 \begin{figure}[tb]
 \begin{center}
\epsfig{figure=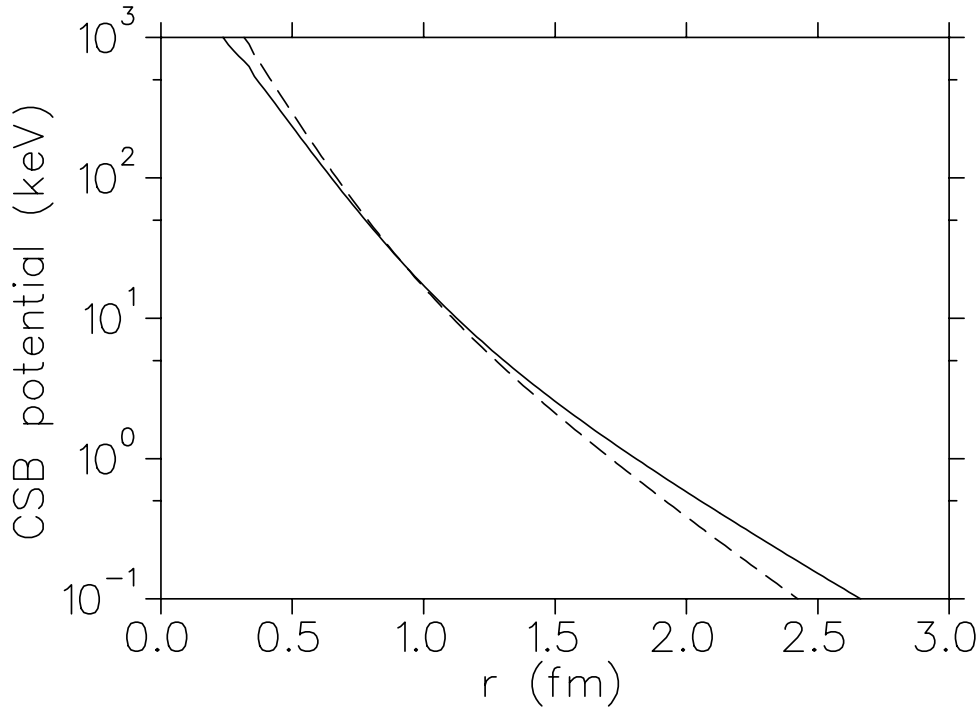,height=13cm,angle=90}
\end{center}
\vspace{-.5cm}
\caption{The coordinate space CSB potentials $V_N(r)$ (solid)
and $V_\Delta (r)$ (dashed) obtained from fits of 
Fig. \protect{\ref{vq}}.}
 \label{vr}
 \end{figure}

The charge symmetric interaction between the nucleons is taken
to be the phenomenological Reid soft core potential \cite{reid}.
This is then also supplemented with explicit excitation of
$N\Delta$ intermediate states by the coupled channels method
\cite{tpe}. No other CSB effects are included in the present
calculation except Eqs. (2-3) (Figs. 1b,c).

The results for the effective range parameter differences
$\Delta a= a_{pp} -a_{nn}$ and $\Delta r_0 = r_{0,pp}-r_{0,nn}$
in the low energy expansion
 $p\, \cot \delta_0 \approx -1/a + \frac 1 2 r_0 p^2$
are presented in Table I. It can be seen that this 
mechanism with the above strength and the monopole
form factor mass $\Lambda = 1\,$GeV gives quite a considerable
contribution to $\Delta a $, about one third
of its experimental value and of the same sign (i.e. the $nn$
interaction is the more attractive of the two). The fraction
is even larger, if one considers that perhaps 0.4 fm in $\Delta
a$ may be attributed simply to different kinetic energies
arising from the $np$ mass difference \cite{henley}.
One half of the calculated effect here
comes from $V_N$ and the other half from the $\Delta$ contribution
$V_\Delta$. 

The column labeled $\Delta E$ is the contribution to the
$^3$H-$^3$He binding energy difference using the
simple prescription
\begin{equation}
\Delta E_{\rm GS} = (40\,\Delta a + 1600 \,\Delta r_0)\,
{\rm keV/fm}
\end{equation}
obtained by Gibson and Stephenson for separable potentials
\cite{gs}. This is likely an overestimate but gives an idea
of the order of magnitude of the effect \cite{cn}. Here the
relevant
empirical result is $\Delta E_{\rm expt} \approx 76\pm 24$ keV
after removing the "trivial" Coulomb repulsion
and the effect of the $np$ mass difference in the kinetic
energy.

\begin{table}[tb]
\begin{tabular}{lccc}
 Model &  $\Delta a$ (fm)  &
$\Delta r_0 $ (fm) &
$\Delta E$(keV) \\
\hline
Reid SC, $N\! N$, only Fig. 1b & 0.28 & 0.006 & 20 \\
Reid SC, $N\! N\! +\! N\!\Delta$, Figs. 1b,c &
0.55 & 0.012 & 41 \\
Reid SC, Figs. 1b,c, dipole ff & 0.40 & 0.009 & 30 \\
 Coupled channels  & 0.50 & 0.010 & 36 \\
Coupled channels, dipole ff & 0.37 & 0.007 & 27 \\
Experiment \protect{\cite{physrep}} & $1.5\pm 0.5$ &
$0.10\pm 0.12$ & $76\pm 24$ \protect{\cite{cn}} \\
\hline & & & \\
\end{tabular}
\caption{CSB effective range parameters
and $^3$H-$^3$He binding energy differences
for various models described in the text.}
\protect\vspace*{3mm}
\end{table}

For model dependence one can vary the form factors. With
softer form factors one
normally expects smaller results. On line 3
the form factor has been taken to be of the dipole form
with the same cut-off mass (or as well monopole
vertices and a dipole formfactor in $\pi N$ scattering).
The result is about 25\% smaller as might be expected.

A more interesting and more fundamental comparison is to a
phase-equivalent
coupled channels calculation with explicit $N\Delta$
intermediate states included in the charge symmetric scattering.
Details of the $NN \leftrightarrow N\Delta$
transition potential  including both $\pi$ and $\rho$ exchanges
are given in Ref. \cite{tpe}. The diagonal $^1S_0$ Reid soft core
potential must be adjusted by a repulsion of
$381\, e^{-3\mu r}/(\mu r)$ MeV to refit the phase shift from
the coupled channels with the original at $E_{\rm lab} =2$ MeV.
By unitarity, the $NN$ wave function should
be depleted at short distances and consequently the
mechanisms 1b,c somewhat suppressed. This is, indeed, the case as
seen on lines 4--5 of Table I, but the decrease is not very
large. Since the $N\Delta$ excitation must
be an essential part of isospin one $NN$ scattering, this may
be considered as the most realistic estimate.

In principle the presence of the $\Delta$ makes new diagrams
possible, e.g. those with one or both pion-baryon vertices
being $\pi\Delta\Delta$ or pion rescattering off the $\Delta$.
The knowledge of these is much inferior to $\pi NN$ or
$\pi N\Delta$. These mechanisms are also of higher order
and are not discussed in this work. However, one could note
that also the above unitarity depletion effect is of
higher order in this sense, so that conservatively one can
only say that the effect of coupling to the $N\Delta$
intermediate states is only of the order of 10\% in the CSB
observables.

The above obtained results appear to indicate that CSB
pion rescattering could be potentially an important effect
also in elastic $NN$ scattering as it was in $np\rightarrow
d\pi^0$. However, we have not yet considered another isospin
violating term in the effective Lagrangian\cite{kolck,weinberg}
\begin{equation}
 {\cal{L}} =
 \frac{\bar\delta m_{N}}{2}  \left(N^\dagger \tau_{0}N
 +\frac{2}{DF_{\pi}^{2}}  N^\dagger (\phi_{0}
\boldphi\cdot\boldtau -\boldphi\cdot\boldphi\,\tau_0)N \right) ,
\end{equation}
of electromagnetic origin. Here $\bar\delta m_{N}$ is the
electromagnetic contribution to the $np$ mass difference,
typically estimated to be of the order of -1 or -2 MeV.
In Ref. \cite{knm} this gave a similar contribution as
Eq. 1 and the strength parameter changed there
$\delta m_N \rightarrow \delta m_N - \bar\delta m_N /2$. 
With $\bar\delta m_N$ negative this increased the effect.
However, in the present case of $NN$ scattering the effect
turns out to be the change 
$\delta m_N \rightarrow \delta m_N + 2 \bar\delta m_N $.
Now the two mass difference terms tend to cancel each other
and the above results should be scaled accordingly by a
factor $(\delta m_N + 
2 \bar\delta m_N ) / (2.4\, {\rm MeV})$ shown as a function
of $\bar\delta m_N$ in Fig. 4.
 It can be seen that for the most reasonable
range of $|\bar\delta m_N|$ between 0.5 and 1.5 MeV \cite{knm}
the strength of the CSB potentials decreases into a fraction 
of the original. For example, using $\bar\delta m_N = -0.76
\pm 0.30$ MeV from the Cottingham formula \cite{gl} 
yielding the strength 2.4 MeV for Ref. \cite{knm}, the final 
results here would be only a quarter of the results in Table I.
This means that if the present situation of understanding
the $pp$ vs. $nn$ difference 
(in particular $\Delta a$) is considered satisfactory, this
understanding is not significantly disturbed by the present
mechanism even if it is large in pion production in
$np \rightarrow d\pi^0$.

\begin{figure}[tb]
\begin{center}
\epsfig{figure=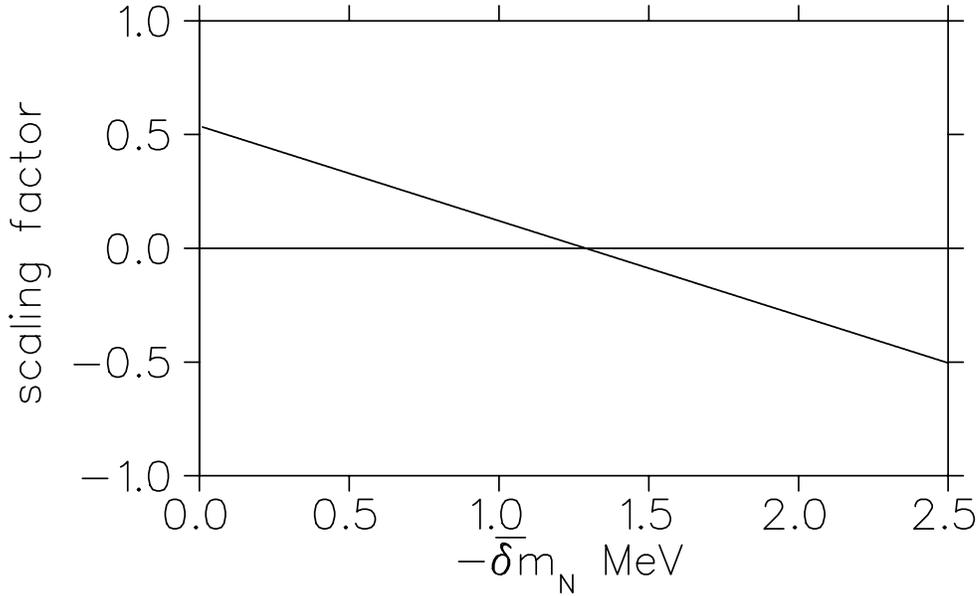,height=13cm,angle=90}
\end{center}
\vspace{-.5cm}
\caption{The scaling factor needed for consistency with the
$np$ mass difference and its electromagnetic part.} 
\label{scale}
\end{figure}

In summary, a new QCD-based $\pi N$ rescattering contribution
has been incorporated in CSB elastic $NN$
scattering using effective field theory. This is potentially
a strong effect as was seen for CSB in pion production.
However, contrary to Ref. \cite{knm}
in this class III interaction the quark mass difference 
and electromagnetic mass contributions tend to cancel, so that
the effect in e.g. $\Delta a$ actually becomes
 rather small. Thus even
the large CSB contribution found in $np \rightarrow d \pi^0$
can be accommodated without compromising the
 understanding of $pp$ vs. $nn$ differences.

\end{document}